\documentclass[aps,prb,superscriptaddress,twocolumn]{revtex4-1}
\usepackage{graphics}
\usepackage{graphicx} % Include figure files
\usepackage{dcolumn}% Align table columns on decimal point
\usepackage{bm}% bold math

\begin{document}

% Use the \preprint command to place your local institutional report
% number in the upper righthand corner of the title page in preprint mode.
% Multiple \preprint commands are allowed.
% Use the 'preprintnumbers' class option to override journal defaults
% to display numbers if necessary
%\preprint{}

%Title of paper
\title{Magnetic structure determination of Ca$_3$LiOsO$_6$ using neutron and x-ray scattering}

% repeat the \author .. \affiliation  etc. as needed
% \email, \thanks, \homepage, \altaffiliation all apply to the current
% author. Explanatory text should go in the []'s, actual e-mail
% address or url should go in the {}'s for \email and \homepage.
% Please use the appropriate macro foreach each type of information

% \affiliation command applies to all authors since the last
% \affiliation command. The \affiliation command should follow the
% other information
% \affiliation can be followed by \email, \homepage, \thanks as well.
%\author{}
%\email[]{Your e-mail address}
%\homepage[]{Your web page}
%\thanks{}
%\altaffiliation{}
%\affiliation{}

\author{S.~Calder}
\email{caldersa@ornl.gov}
\affiliation{Quantum Condensed Matter Division, Oak Ridge National Laboratory, Oak Ridge, Tennessee 37831, USA.}

\author{M.~D.~Lumsden}
\affiliation{Quantum Condensed Matter Division, Oak Ridge National Laboratory, Oak Ridge, Tennessee 37831, USA.}

\author{V.~O.~Garlea}
\affiliation{Quantum Condensed Matter Division, Oak Ridge National Laboratory, Oak Ridge, Tennessee 37831, USA.}

\author{J.-W.~Kim}
\affiliation{Advanced Photon Source, Argonne National Laboratory, Argonne, Illinois 60439, USA.}

\author{Y.~G.~Shi}
\affiliation{Institute of Physics, Chinese Academy of Sciences, 100190 Beijing, China.}
\affiliation{Superconducting Properties Unit, National Institute for Materials Science, 1-1 Namiki, Tsukuba, Ibaraki 305-0044, Japan.}

\author{H.~L.~Feng}
\affiliation{Superconducting Properties Unit, National Institute for Materials Science, 1-1 Namiki, Tsukuba, Ibaraki 305-0044, Japan.}
\affiliation{Graduate School of Chemical Sciences and Engineering, Hokkaido University, Sapporo, Hokkaido 060-0810, Japan.}

\author{K.~Yamaura}
\affiliation{Superconducting Properties Unit, National Institute for Materials Science, 1-1 Namiki, Tsukuba, Ibaraki 305-0044, Japan.}
\affiliation{Graduate School of Chemical Sciences and Engineering, Hokkaido University, Sapporo, Hokkaido 060-0810, Japan.}

\author{A.~D.~Christianson}
\affiliation{Quantum Condensed Matter Division, Oak Ridge National Laboratory, Oak Ridge, Tennessee 37831, USA.}

%Collaboration name if desired (requires use of superscriptaddress
%option in \documentclass). \noaffiliation is required (may also be
%used with the \author command).
%\collaboration can be followed by \email, \homepage, \thanks as well.
%\collaboration{}
%\noaffiliation

\date{\today}

\begin{abstract}
We present a neutron and x-ray scattering investigation of Ca$_3$LiOsO$_6$, a material predicted to host magnetic ordering solely through an extended superexchange pathway involving two anions, an interaction mechanism that has undergone relatively little investigation. This contrasts with the ubiquitous superexchange interaction mechanism involving a single anion that has well defined and long standing rules. Despite the apparent 1D nature and triangular units of magnetic osmium ions the onset of magnetic correlations has been observed at a high temperature of 117 K in bulk measurements. We experimentally determine the magnetically ordered structure and show it to be long range and three dimensional. Our results support the model of extended superexchange interaction. 
\end{abstract}

% insert suggested PACS numbers in braces on next line
% insert suggested keywords - APS authors don't need to do this
%\keywords{}

%\maketitle must follow title, authors, abstract, \pacs, and \keywords
\maketitle

\section{\label{sec:Introduction}Introduction}

The investigation of $5d$ transition metal oxides (TMO) has resulted in the observation of a variety of novel properties due to the competition of spin-orbit coupling (SOC), on-site Coulomb interaction and crystal field splitting that are all of comparable strength. This contrasts with the much more studied $3d$ TMO in which spin-orbit coupling is generally only a small perturbation. On the other hand the radius of the electronic wavefunction is extended in $5d$ systems compared to $3d$ TMO, resulting in increased itinerant properties. Much of the recent experimental and theoretical focus on $5d$ TMO has concentrated on iridates due to the observance of a so called $J_{\rm eff} = 1/2$ Mott spin-orbit insulating state. This state arises due to the SOC splitting of the $t_{2g}$ manifold as a consequence of the $5d^5$ electron configuration. Consequently even small on-site Coulomb interactions can result in insulating behavior \cite{KimScience}. Iridates have been found to host topological insulating states \cite{NaturePesin}, Weyl semi-metal \cite{PhysRevB.83.205101} and potentially the Kitaev model \cite{PhysRevLett.105.027204}.

\begin{figure}[b]
     \centering                      
\includegraphics[width=0.49\columnwidth]{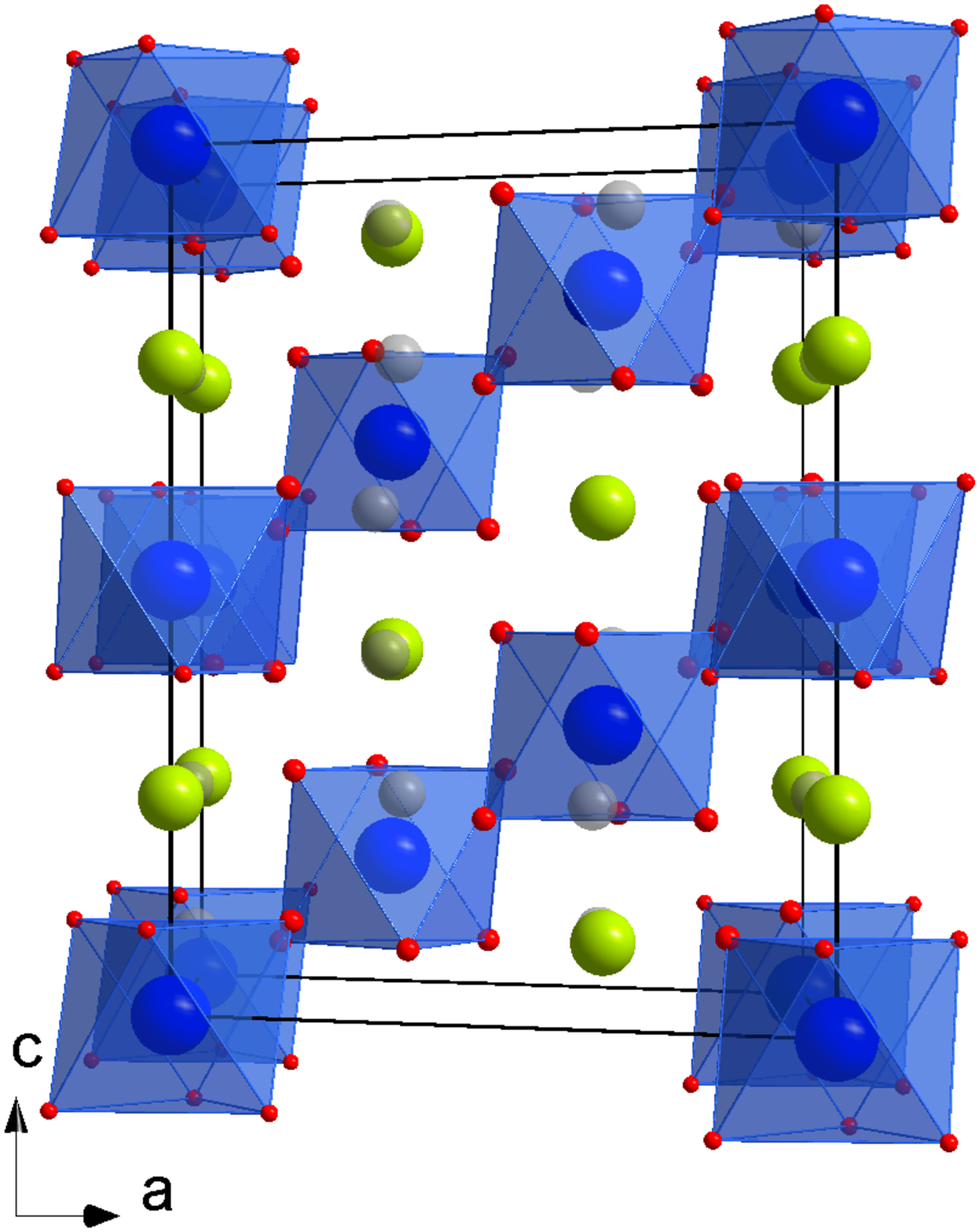}
                       \includegraphics[width=0.49\columnwidth]{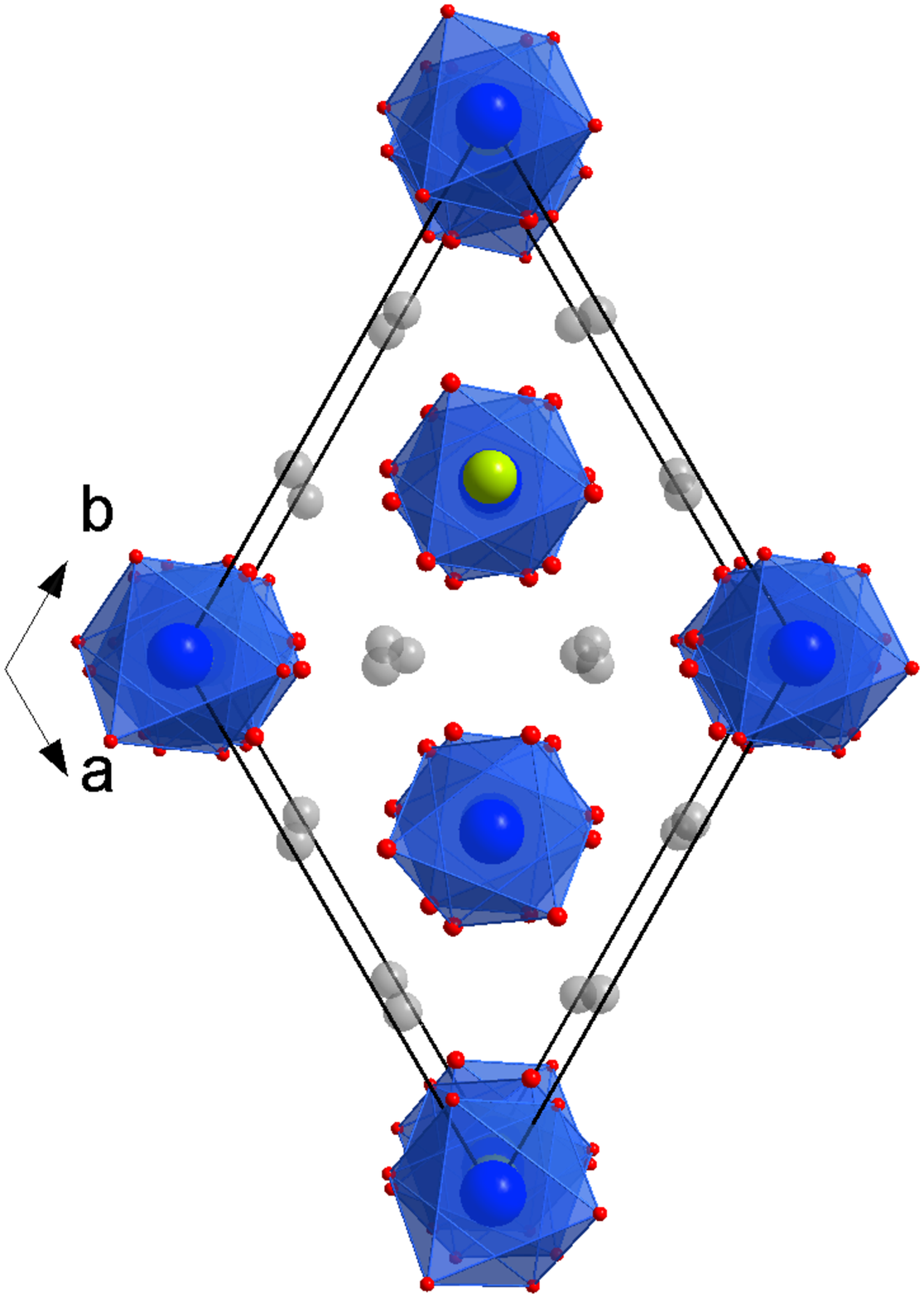} 
 \caption{\label{FigureXtalStruct} Ca$_3$LiOsO$_6$ forms a hexagonal crystal structure with space group $R\bar3c$. Magnetic Os$^{5+}$ ions (blue spheres) are surrounded by six oxygen anions (red spheres). These octahedra are separated along the $c$-axis by Li$^+$ ions. Ca$^{2+}$ ions (grey spheres) provide the charge balance.}
\end{figure}

Systems containing the neighboring osmium ion have also showed interesting properties. For example NaOsO$_3$ undergoes a Slater metal-insulator transition \cite{NaOsO3Calder} and KOs$_2$O$_6$ is an unconventional superconductor \cite{KOs2O6SC}. We focus on an additional compound of note, Ca$_3$LiOsO$_6$, that was grown and characterized for the first time by Shi {\it et al.}\cite{ShiJACS}. Ca$_3$LiOsO$_6$ forms the K$_4$CdCl$_6$-type crystal structure (hexagonal space group $R\bar3c$) in which apparent 1D chains of Os ions along the $c$-axis are separated by Li ions in a frustrated geometry, as shown in Fig \ref{FigureXtalStruct}.  However, the high magnetic ordering temperature reported of 117 K is not compatible with this quasi-1D frustrated picture \cite{ShiJACS}. Instead the authors presented a model for Ca$_3$LiOsO$_6$ containing extended superexchange magnetic interactions of Os-O-O-Os to account for the high magnetic ordering temperature. Shortly after a theoretical investigation supported this model and considered possible magnetic exchange interactions \cite{KanInorChem}. Whilst magnetic interactions mediated by a single anion are common with long standing and well defined ÒGoodenough-Kanamori rulesÓ \cite{GoodenoughSE}, the extended superexchange interaction through two anions has undergone relatively little investigation.

We have performed a neutron and magnetic x-ray scattering investigation to determine the magnetic structure of Ca$_3$LiOsO$_6$.  The related $4d$ material Ca$_3$LiRuO$_6$ also undergoes a magnetic transition at a similar temperature, suggestive that the electronic configuration of $d^3$ is significant and plays a role in the magnetic ordering \cite{Darriet1997139}.  For Ca$_3$LiOsO$_6$ we find a magnetic structure in which both inter and intra-chain magnetic interactions are present. The magnetic model is 3D and contains no geometric frustration, making it compatible with the high magnetic ordering temperature observed and the low ratio of ${\rm \Theta_W/T_N}$ from susceptibility \cite{ShiJACS}. We present both powder and single crystal results that confirm the magnetic ordering temperature and are consistent with a magnetic structure that involves the extended superexchange interaction. Ca$_3$LiOsO$_6$, along with Ca$_3$LiRuO$_6$, appear to be ideal candidates to investigate the extended superexhange interaction.  Unlike the iridates discussed above in which the $5d^5$ electronic configuration leads to the degeneracy of the $t_{2g}$ level being broken by spin-orbit coupling, the $5d^3$ configuration is expected to remain degenerate, even for the case of large SOC \cite{ChenPRBd2}. We consider the $t_{2g}$ configuration through an interpretation of our neutron and x-ray results.

\section{\label{sec:ExptMethods}Experimental Methods}

Single crystal and polycrystalline samples of Ca$_3$LiOsO$_6$ were prepared as described in Ref.~\onlinecite{ShiJACS}. Neutron powder diffraction (NPD) measurements were performed at the High Flux Isotope Reactor (HFIR) at Oak Ridge National Laboratory (ORNL) using beamline HB-2A. Measurements were performed with both $\lambda = 1.54$ ${\rm \AA}$ and $\lambda = 2.41$ ${\rm \AA}$ on a 5 gram sample. The shorter wavelength gives a greater intensity and higher $Q$ coverage that we utilized to investigate the crystal structure through the magnetic phase transition from 150 K to 30 K. The longer wavelength gives lower $|Q|$ coverage and greater resolution that we employed to investigate the magnetic structure at 4 K, with a comparable measurement at 150 K. The NPD data was investigated using the rietveld refinement program Fullprof and the magnetic structural representational analysis was performed using SARA$h$ \cite{sarahwills}. 

The triple-axis instrument HB1 at HFIR was used in elastic mode with a wavelength of 2.46 ${\rm \AA}$ to investigate the magnetic order parameter at the (101) magnetic reflection on the same powder sample. Various measurements were taken from 10 K through the magnetic transition temperature of $\sim$$117$ K. The integrated intensity of the scattering around the magnetic reflection was calculated to determine the magnetic ordering temperature. 

A single crystal investigation on a crystal of approximate size 0.1$\times$0.1$\times$0.05 mm was performed at the Advanced Photon Source (APS) at beamline 6-ID-B using magnetic resonant X-ray scattering (MRXS). We carried out measurements at both the L2 and L3 resonant edges of osmium that correspond to 12.393 keV and 10.878 keV, respectively. Graphite was used as the polarization analyzer crystal at the (0,0,10) and (0,0,8) reflections on the L2 and L3 edges, respectively, to achieve a scattering angle close to 90$^\circ$. For the L3 edge the scattering angle was 86$^{\circ}$ and for L2 the scattering angle was 94$^{\circ}$.  Measurements were taken at several reflections to investigate possible magnetic structures, with an analysis of the photon polarization in $\sigma$-$\sigma$ and $\sigma$-$\pi$ allowing a distinction between magnetic and charge scattering. To account for absorption, energy scans were performed without the analyzer and with the detector away from any Bragg peaks through both absorption energies.

\section{\label{sec:ExptRresults}Results and Discussion}

\subsection{\label{sec:crystalstruct}Neutron diffraction crystal structure investigation}

\begin{figure}[b]
     \centering
 \includegraphics[width=0.9\columnwidth]{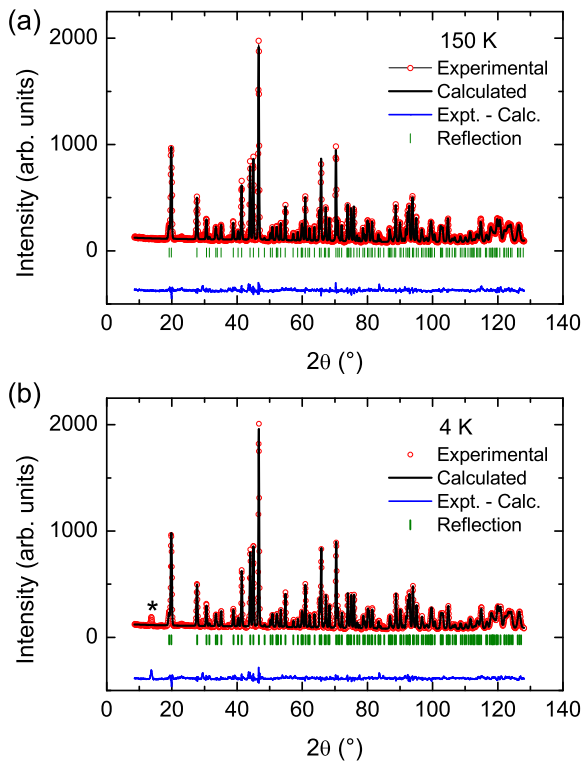}          
 \caption{\label{Figure2Calder}Neutron powder diffraction results at (a) 150 K and (b) 4 K with $\lambda = 1.54$ $\rm \AA$ measured on HB-2A at HFIR. The crystal structure symmetry remains unchanged between high and low temperature, however an additional magnetic reflection indicative of magnetic order is observed at low temperature, indicated by the asterix.}
\end{figure}

\begin{figure}[htb]
     \centering
          \includegraphics[width=1.0\columnwidth]{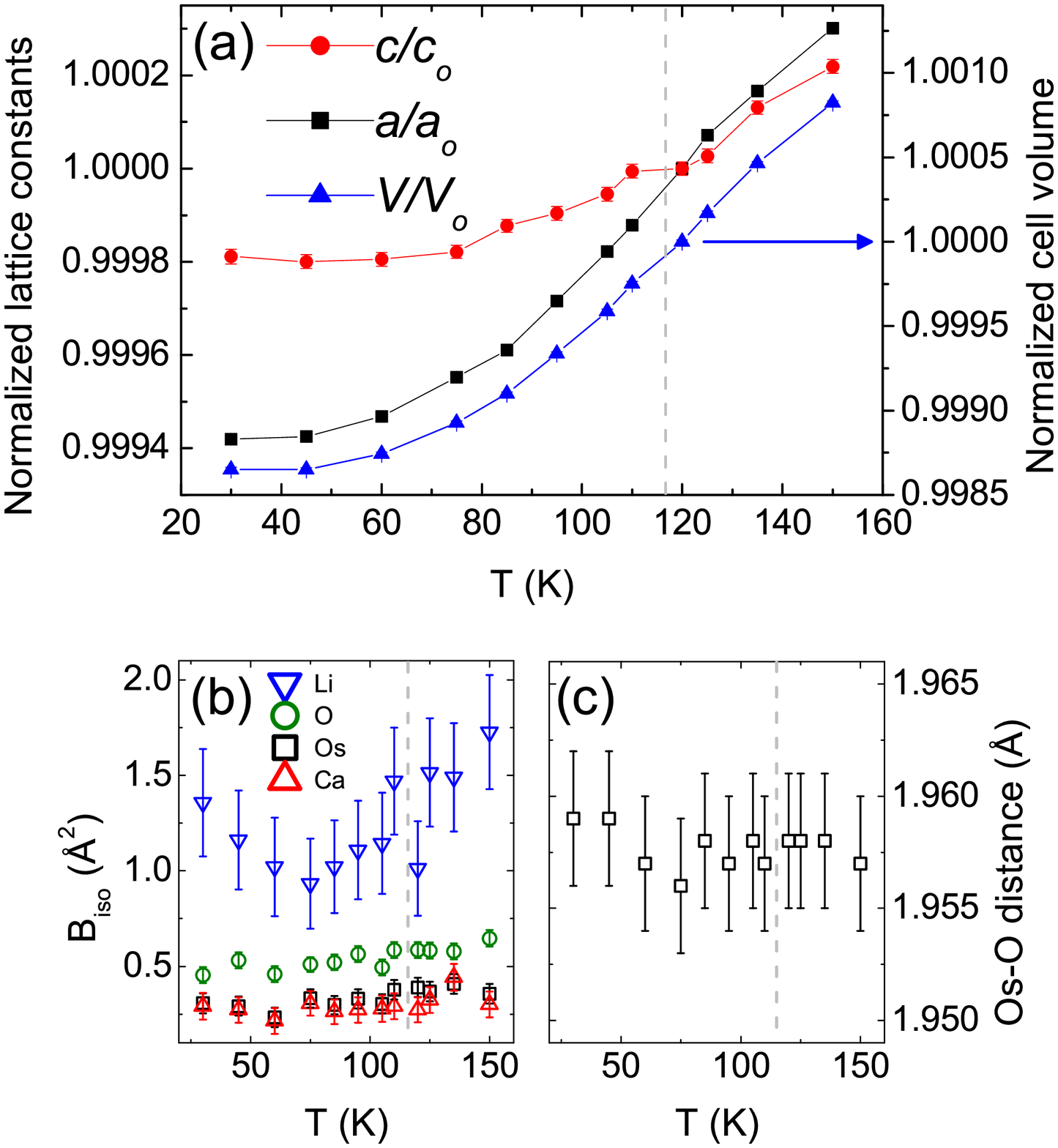}  
 \caption{\label{FigureUnitCellVar} (a) Normalized unit cell parameters for Ca$_3$LiOsO$_6$. The values used to normalize the parameters correspond to the lattice parameters at 120 K of $V_o=798.1329$ $\rm \AA$, $a_o = 9.2545$ $\rm \AA$ and $c_o=10.7607$ $\rm \AA$. (b) Thermal parameters, $\rm B_{iso}$ provide an indication for the stability of the structural model. (c) Os-O bond distance in the octahedra. The grey dashed line indicates the magnetic transition temperature.}
\end{figure}

We investigated the crystal structure temperature variation of Ca$_3$LiOsO$_6$ using NPD from 150 K to 4 K, through the magnetic anomaly observed around 117 K \cite{ShiJACS}. The crystal structure temperature dependence has been previously followed down to low temperature with x-ray diffraction \cite{ShiJACS}, however the insensitivity of XRD to oxygen ions necessitates further studies using neutron scattering to understand the overall crystal structural behavior. Figure \ref{Figure2Calder} and Table \ref{FPtableperams} show NPD diffraction results at 4 K and 150 K that are consistent with previously reported XRD \cite{ShiJACS}. There is no evidence of a crystal structure symmetry change between 150 K and 4 K. The thermal parameters B$_{\rm iso}$, shown in Fig.~\ref{FigureUnitCellVar}, give an indication of the stability of the structural model and remain consistent for all temperatures measured. At low temperature additional scattering is observed at commensurate positions that are indicative of magnetic order, we defer discussion of magnetic ordering until section \ref{sec:magnstruct}.  

\begin{table}[tb]
\caption{\label{FPtableperams}Refined crystal structure parameters from Fullprof for (a) 150 K and (b) 4 K.}
\begin{flushleft}
(a) 150 K
\end{flushleft}
\begin{tabular}{l l}
Space group& $R\bar3c$ \\
$a$ &$9.257(1)$ $\rm \AA$ \\
$c$ & $10.763(1)$ $\rm \AA$\\
Cell volume  &798.86(1) $\rm \AA^{3}$\\
$\chi^2$ &  4.30\\
R$_{\rm wp}$ & 5.99$\%$.\\
\end{tabular} 
\vspace{0.15cm} \\
		\begin{tabular}{c c c c c c c c}
\hline 
Atom  &   site & $x$ & $y$& $z$ & $B_{iso}$ ($\rm \AA^2$)  \\ \hline
Os  & $6b$ &  0 & 0 &  0  & 0.366(4) \\
Ca  & $18e$ &  0.6460(3) &0 & 0.25     & 0.333(5)\\
O & $36f$ &  0.0281(14)  &   0.8438(1) &       0.3942(2)    & 0.610(3)\\    
Li  & $6a$ & 0 &   0 &    0.25 &   1.54(2)\\  
\hline
\end{tabular}
\vspace{0.15cm} \\
\begin{flushleft}
(a) 4 K
\end{flushleft}
\begin{tabular}{l l}
			Space group& $R\bar3c$ \\
			  $a$ &$9.249(1)$ $\rm \AA$ \\
			  $c$ & $10.758(1)$ $\rm \AA$\\
			  Cell volume  &797.02(1) $\rm \AA^{3}$\\
			$\chi^2$ &  5.85\\
			R$_{\rm wp}$ & 5.83$\%$. \\
\end{tabular}
		\begin{tabular}{c c c c c c c c}
\hline 
Atom  &   site & $x$ & $y$& $z$ & $B_{iso}$ ($\rm \AA^2$)  \\ \hline
Os  & $6b$ &  0 & 0 &  0  & 0.293(3) \\
Ca  & $18e$ &  0.6458(3) &0 & 0.25     & 0.209(4)\\
O & $36f$ &  0.0279(2)  &   0.8437(2) &       0.3940(2)    & 0.483(3)\\    
Li  & $6a$ & 0 &   0 &    0.25 &   1.14(2)\\  
\hline
\end{tabular}
\vspace{0.3cm} \\
\end{table}

To investigate the crystal structural behavior through the magnetic transition of 117 K we performed NPD from 150 K to 30 K and modeled the results with Fullprof to obtain unit cell and atomic parameters. The variation of the unit cell is shown in Fig.~\ref{FigureUnitCellVar}. The unit cell volume shows no indication of a structural anomaly through the magnetic transition, with the lattice compression rate decreasing at low temperatures. Considering the $a$ and $c$ lattice constants we find that the $a$-axis follows the trend of the overall unit cell. The rate of thermal contraction for the c-axis  is much reduced. However, our NPD results are not conclusive to assign the possibility of a structurally driven magnetic transition and any anomaly is much less pronounced than from XRD \cite{ShiJACS}.

The crystal structure of Ca$_3$LiOsO$_6$ has each osmium ion surrounded by six oxygen ions in an octahedral environment, see Fig.~\ref{FigureXtalStruct}. The octahedra remain invariant with $a$ or $b$ axis translations, but their orientation alternate along the $c$-axis.  In a model of extended magnetic superexchange interaction the octahedra form the Os-O-O-Os exchange pathway. Figure \ref{FigureUnitCellVar} shows the temperature variation of the Os-O distance obtained from NPD. There is virtually no deviation with temperature. The electronic configuration of the Os$^{5+}$ ion is $5d^3$, with the three electrons therefore expected to each occupy the crystal field split degenerate $t_{2g}$ orbitals $d_{xy}$, $d_{yz}$ and $d_{zx}$. Ref.~\onlinecite{ShiJACS} discusses the possibility of a trigonal CEF breaking of the $t_{2g}$ degeneracy.  If this degeneracy were broken it would be expected that this would be reflected in a deviation of the Os-O bonds within the octahedra, however this is not observed. Similarly any Jahn-Teller distortions are precluded by the lack of variation in the Os-O distance. Therefore our results show no indication that the $t_{2g}$ degeneracy that results due to the crystal field splitting of octahedrally coordinated oxygens anions around the magnetic ion being further split by spin-orbit interactions, as occurs in certain irridates.\\
%{\it The point group is hexagonal, therefore check carefully that there is no unexpected CEF splitting. However local symmetry for Os ion is cubic.}

\subsection{\label{sec:magnstruct}Magnetic structure}

\subsubsection{Neutron powder diffraction}

\begin{figure}[tb]
     \centering
                       \includegraphics[width=1.0\columnwidth]{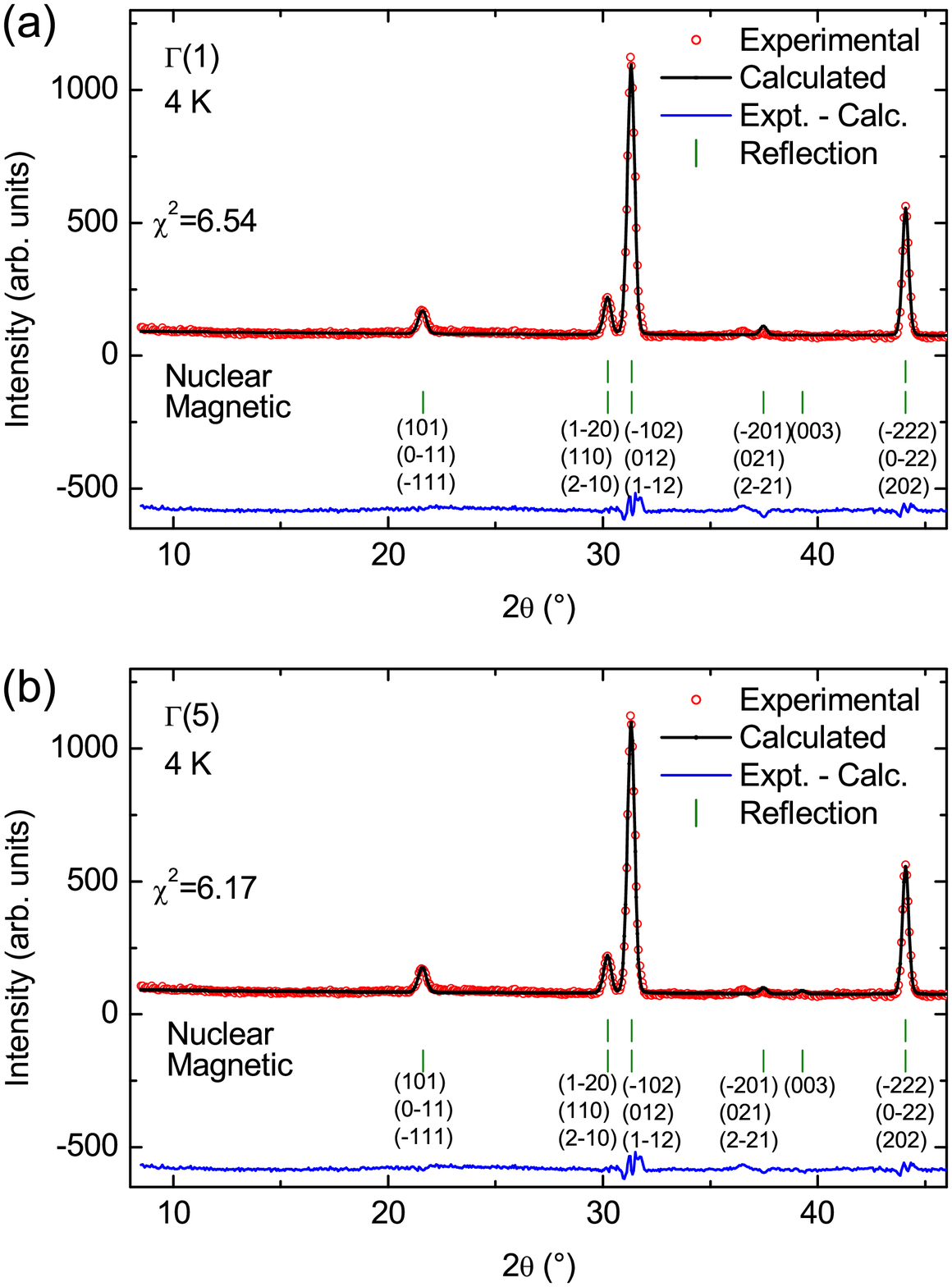}
		\includegraphics[width=0.49\columnwidth]{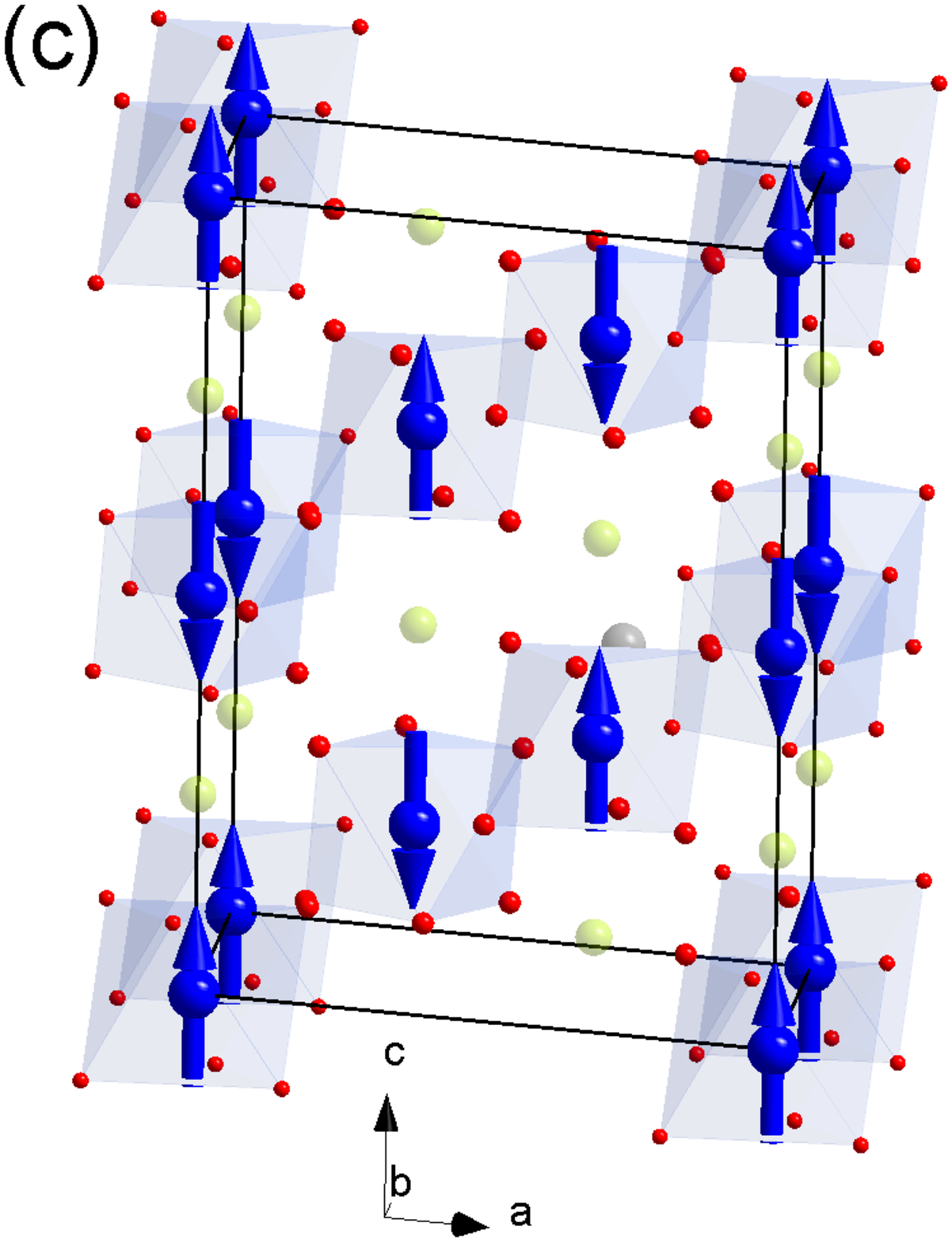}
                       \includegraphics[width=0.49\columnwidth]{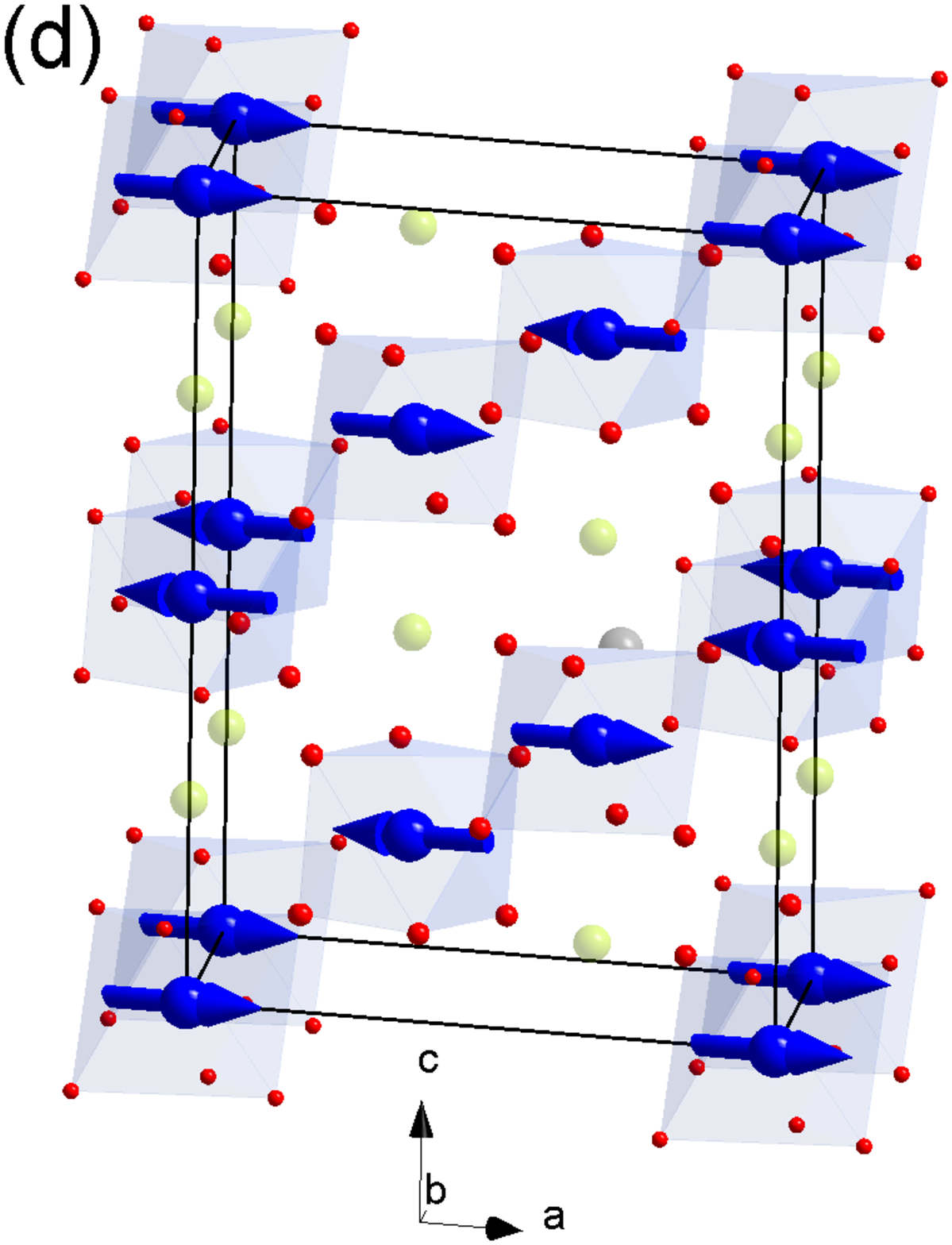}
 \caption{\label{FigureMagFP_struct}NPD data with $\lambda = 2.41$ $\rm \AA$ at 4 K modeled with two possible magnetic structures from the representational analysis, labelled (a) $\Gamma(1)$ and (b) $\Gamma(5)$, following the numbering scheme of Kovalev. The corresponding magnetic spin structure is shown for (c) $\Gamma(1)$ and (d) $\Gamma(5)$. The schematic shows 180$^{\circ}$ between spins, however additional angles are possible in the basal plane as described in the text.}
\end{figure}

The NPD measurements showed additional scattering at commensurate  positions  below $\sim$120 K indicative of magnetic ordering (see Fig.~\ref{Figure2Calder}).  To determine the nature of the magnetic order we implemented representational analysis \cite{sarahwills}. For a second order transition, Landau theory states that the symmetry properties of the magnetic structure are described by only one irreducible representation (IR). We performed our analysis with a propagation vector with various non-zero values, e.g.~(001) etc, however the models were never found to be compatible with the observed magnetic scattering and therefore we concentrate solely on a ${\bf k} = (000)$ magnetic structure. For the $R\bar3c$ crystal structure with the magnetic moment on the Os ion and commensurate propagation vector, ${\bf k} = (000)$, there are three possible IRs. Table \ref{basis_vector_table_1} lists the IR and corresponding basis vectors, $\psi$.  The IR correspond to $\Gamma(1)$, $\Gamma(3)$ and $\Gamma(5)$ (following the numbering scheme of Kovalev  \cite{Kovalev}). $\Gamma(3)$ could be readily discarded as not giving scattering at the correct reflections for the magnetic ordering. Figure \ref{FigureMagFP_struct} shows the refined model for the $\Gamma(1)$ and $\Gamma(5)$ IRs. Refining both models to fit the experimental scattering does not produce conclusively different results to allow for the definition of a unique magnetic structure, despite relatively large counting times performed during the NPD. 

\begin{table}[tb]
\begin{tabular}{ccc|rrrrrr}
  IR  &  BV  &  Atom & \multicolumn{6}{c}{BV components}\\
      &      &             &$m_{\|a}$ & $m_{\|b}$ & $m_{\|c}$ &$im_{\|a}$ & $im_{\|b}$ & $im_{\|c}$ \\
\hline
$\Gamma_{1}$ & $\psi_{1}$ &      1 &      0 &      0 &      1 &      0 &      0 &      0  \\
             &              &      2 &      0 &      0 &     -1 &      0 &      0 &      0  \\
$\Gamma_{3}$ & $\psi_{2}$ &      1 &      0 &      0 &      1 &      0 &      0 &      0  \\
             &              &      2 &      0 &      0 &      1 &      0 &      0 &      0  \\
$\Gamma_{5}$ & $\psi_{3}$ &      1 &      6 &      0 &      0 &      0 &      0 &      0  \\
             &              &      2 &      0 &      0 &      0 &      0 &      0 &      0  \\
             & $\psi_{4}$ &      1 &      0 &      0 &      0 &      0 &      0 &      0  \\
             &              &      2 &     -1 &     -1 &      0 &      0 &      0 &      0  \\
             & $\psi_{5}$ &      1 &      0 &      0 &      0 &      0 &      0 &      0  \\
             &              &      2 & -1 &  1 &      0 &      0 &      0 &      0  \\
             & $\psi_{6}$ &      1 & -1 & -1 &      0 &      0 &      0 &      0  \\
             &              &      2 &      0 &      0 &      0 &      0 &      0 &      0  \\
\end{tabular}
\caption{Basis vectors for the space group R-3c:H with 
${\bf k}=(0,0,0)$.The decomposition of the magnetic representation for the Os site $( 0,0,0)$ is $\Gamma_{Mag}=1\Gamma_{1}^{1}+0\Gamma_{2}^{1}+1\Gamma_{3}^{1}+0\Gamma_{4}^{1}+2\Gamma_{5}^{2}+0\Gamma_{6}^{2}$. The atoms of the nonprimitive basis are defined according to 1: $( 0,0,0)$, 2: $( 0,0,\frac{1}{2})$.}
\label{basis_vector_table_1}
\end{table}

Although the scattering is well reproduced by both models, there is a clear distinction between the magnetic structure they represent. The two candidate magnetic structures are shown schematically in Fig.~\ref{FigureMagFP_struct}(c)-(d). $\Gamma(1)$ has spins AFM aligned along the $c$-axis in 1D chains, for which 3D ordering would be frustrated due to the triangular units between the chains. $\Gamma(5)$ has spins oriented in the $a$-$b$ plane that would require extended superexchange Os-O-O-Os interactions to describe the 3D ordering.

There are certain distinctions between the modeled neutron scattering for each IR. Firstly, for $\Gamma(1)$ there is no scattering at the reflection at $\sim$40$^{\circ}$ that corresponds to (003), whilst there is scattering for the $\Gamma(5)$ model. Secondly, the scattering at the magnetic peaks is slightly better modeled for the $\Gamma(5)$ model compared to the $\Gamma(1)$ model as evidenced by the  $\chi^2$ value being slightly lower for the $\Gamma(5)$ model. As shown in Table \ref{basis_vector_table_1} there is only one BV, $\psi_1$, for  $\Gamma(1)$. Therefore there are only moments allowed oriented in $c$-axis and consequently the only variable is the size of the magnetic moment. The refinement for this $\Gamma(1)$ model gives a magnetic moment of $\sim$2$\mu_B$. Conversely for $\Gamma(5)$ there are four basis vectors, that all only allow moments in the $a$-$b$ plane. In the basal plane any angle between spins is possible. However, placing the spins at an angle of 90$^{\circ}$ or less results in an intensity mismatch on the lowest two observed angular reflections and can be readily discarded as a valid model. In the IR analysis The atoms of the non-primitive basis are defined according to 1: $( 0,0,0)$, 2: $( 0,0,\frac{1}{2})$. These two sites sit at equivalent atomic environments and so we consider the constraint that the magnitude of the spins on both sites should be the same and the orientation restricted to equivalent directions in the hexagonal crystal structure. Using different basis vector values with these constraints produces magnetic models with fixed angle of 120$^{\circ}$ and 180$^{\circ}$ between the spins in the basal plane as physically reasonable models. The modeled intensity is different for different values of BVs at the allowed reflections for $\Gamma(5)$. However, the sharp fall of the form factor dependence for Os$^{5+}$ results in appreciably no discernible difference between the various BV values with angles between spins of 120$^{\circ}$ and 180$^{\circ}$. Consequently from the NPD data for the $\Gamma(5)$ model we are able to define the the magnetic moment as falling in the range 2.0$\mu_B$ to 2.3$\mu_B$, with the spins in the  $a$-$b$ plane.

Both $\Gamma(1)$ and $\Gamma(5)$ models give a reduced magnetic moment from the spin only $S=3/2$ value of 3$\mu_B$ of between 66$\%$ to 76$\%$. This is reduced from that predicted by a Curie-Weiss fit to the susceptibility that produced 98$\%$ of the expected spin only model \cite{ShiJACS}. A reduced moment from that expected by a localized spin model, and even that found from a Curie-Weiss fit, has been observed for various $5d$ TMO systems \cite{PhysRevLett.106.067201}. The extended radius of $5d$ systems would be expected to result in a greater tendency away from a purely localized spin model and consequently a larger degree of covalency and charge fluctuations has been postulated to describe this behavior \cite{PhysRevLett.108.197202}.

\subsubsection{Magnetic resonant X-ray scattering}

To distinguish between the $\Gamma(1)$ and $\Gamma(5)$ magnetic models we extended our investigation to a single crystal of Ca$_3$LiOsO$_6$ using magnetic resonant X-ray scattering (MRXS) at the APS. MRXS has several unique qualities, specifically for our investigation is the ability to measure magnetic scattering using small single crystals of dimensions inaccessible to neutron scattering. Additionally MRXS directly probes electronic excitations within the magnetic ion and therefore allows a consideration of the role of the increased SOC in $5d$ systems, as has been previously shown using MRXS \cite{KimScience}. In general $5d$ systems are particularly well suited to MRXS and show large resonant enhancements at certain well defined energies  \cite{Hill.Ir.Kitaev, KimScience, 0953-8984-15-2-108}. For osmium these resonant edges are labelled L2 and L3 and correspond to energies of 12.393 keV and 10.878 keV, respectively, for Ca$_3$LiOsO$_6$. The L3 absorption corresponds to electronic $2p_{\frac{3}{2}} \rightarrow 5d$ transitions and the L3 edge corresponds to $2p_{\frac{1}{2}} \rightarrow 5d$ electronic transitions.  The crystal environment plays a role in determining the exact resonant energies observed away from that of an isolated single ion, however we note that Ca$_3$LiOsO$_6$ and the recently measured NaOsO$_3$  produced virtually the same resonant energy values and energy peak shapes \cite{NaOsO3Calder}. This suggests a similar local environment and electronic configuration for the Os$^{5+}$ ion in both systems. We observed a strong magnetic resonant enhancement at both L2 and L3 edges for Ca$_3$LiOsO$_6$. This is compatible with a lack of splitting of the $t_{2g}$ degeneracy by spin-orbit coupling, as was observed for NaOsO$_3$. These results contrast with irridates with a $J_{\rm eff} = 1/2$ insulating state in which there is virtually no resonant enhancement at the L2-edge \cite{KimScience}.

\begin{table}[tbp]
\caption{\label{MRXSreflections}Measured reflections of Ca$_3$LiOsO$_6$ using MRXS. Those marked ($^*$) are reflections corresponding to only the  $\Gamma(5)$ model.}
\begin{center}
\begin{tabular}{l c}
\hline
\hline
Reflection & Magnetic \\
\hline
\hline
(0 0 3)$^*$ & yes \\
(1  -1   5) & yes  \\
(2 0 5) & yes \\
(5  0   5) &  yes \\

(1   0   7) &  yes\\
(0   -1   7) &  yes\\
(-1 1 7) &  yes\\
(-1 -2 7) & yes \\
(4   0   7) &  yes\\
(0 0 9)$^*$ & yes \\

(-2 2 11) & yes \\
(-1 0 11) & yes \\
(0 1 11) & yes \\
(1  -1  11) &yes  \\
(0  -2  11) & yes \\

(0 0 15)$^*$& yes \\
\hline
(-1  0   7) &  no scattering \\
(0  1   7) &  no scattering \\
\hline
(1 1 9) & magnetic and charge scattering \\
(1   2  11) &  magnetic and charge scattering \\
\hline
\end{tabular}
\end{center}
\label{default}
\end{table}%

\begin{figure}[htb]
     \centering
                       \includegraphics[width=0.95\columnwidth]{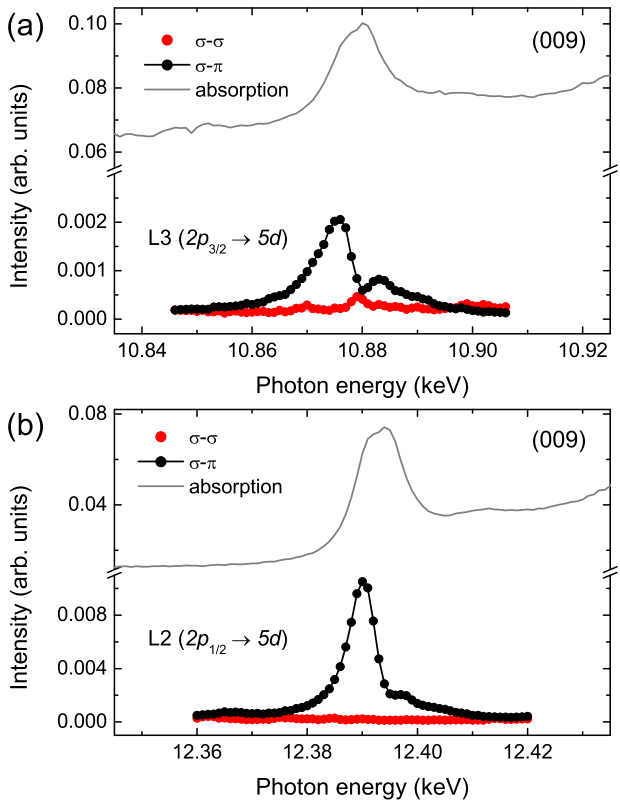}
 \caption{\label{FigureMRXS} MRXS energy dependence of the (009) reflection.  There is a large enhancement observed in the $\sigma$-$\pi$ scattering and a suppression of scattering in the $\sigma$-$\sigma$ channel, as expected for a purely magnetic reflection. A large resonant enhancement was observed at both L2 and L3 edges.}
\end{figure}

We performed a thorough analysis of various magnetic and non-magnetic reflections using MRXS to distinguish between the two candidate magnetic structures from NPD. To determine if the observed scattering was magnetic or non-magnetic (or both) we performed a polarization analysis of the scattered x-ray beam. This exploits the fact that an incident x-ray beam of linearly polarized light is rotated by 90$^{\circ}$ when scattered by magnetic dipoles. Therefore for $\sigma$-$\sigma$ polarization there is no intensity if the scattering is magnetic, whereas $\sigma$-$\pi$ polarization produces a large enhancement around the resonant edges for magnetic scattering. This is demonstrated in Fig.~\ref{FigureMRXS} for the (009) purely magnetic reflection. To confirm the consistency of our distinction between magnetic and non-magnetic reflections we measured all permutations of a select reflection: (1,0,7), (0,-1,7),(-1,1,7),(-1,0,7) and (0,1,7) and found the correct nature of scattering for the crystal and magnetic structures.

Table \ref{MRXSreflections} lists the measured peaks of Ca$_3$LiOsO$_6$ in our investigation and the nature of the scattering. All the magnetic reflections are consistent with that of the $\Gamma(5)$ model, with those marked with an asterix not allowed for the $\Gamma(1)$ model. Therefore taking all our neutron and x-ray results support the $\Gamma(5)$ model, with spins oriented in the $a$-$b$ plane, as being the ordered magnetic structure for Ca$_3$LiOsO$_6$. We present results of MRXS scans through (H,K,L) at the (-2,2,11) magnetic reflection. The scans show peaks at (H,K,L) that are entirely consistent with 3D magnetic order.

\begin{figure}[tb]
     \centering
                       \includegraphics[width=0.9\columnwidth]{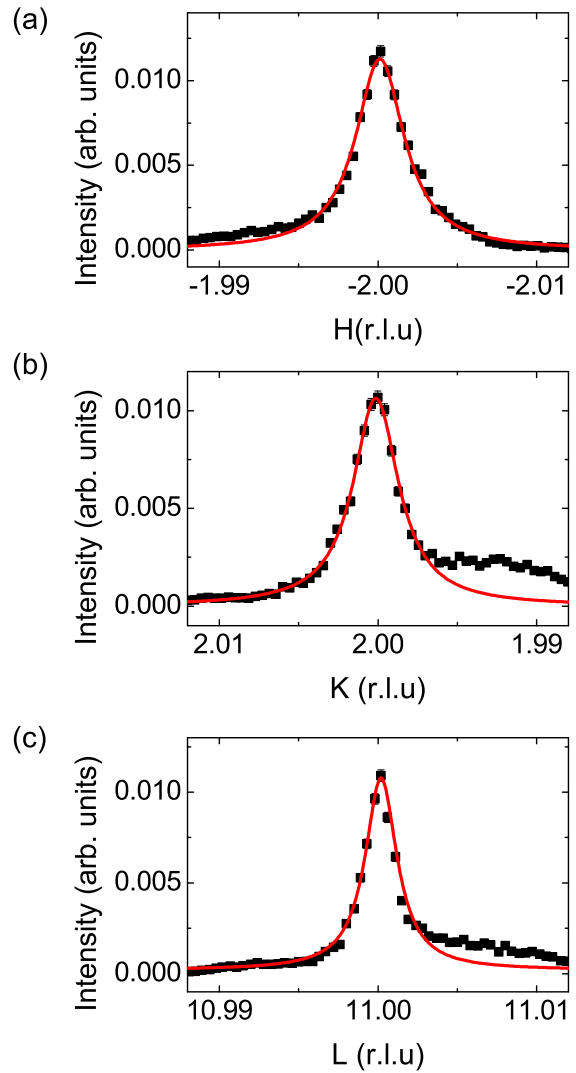}
 \caption{\label{FigureHKL_Xray} MRXS measurement of the magnetic reflection (-2,2,11) showing (H,K,L) scans that support 3D magnetic ordering. (a) H scan at constant K and L, (b) K scan at constant H and L and (b) L scan at constant H and K. The values have been normalized to (-2,2,11).}
\end{figure}

One question remains as to the specific direction of each spin in the $a$-$b$ plane or if indeed there is a specific direction or a random arrangement due to domains. Since rotating the spins equally around the $a$-$b$ plane results in appreciably the same modeled scattering from neutron diffraction this could not be used to distinguish between directions. Additionally we performed azimuthal scans that can give definitive information on the moment direction, however we observed no obvious trends. Regardless of the specific spin direction, the relative arrangement of each spin with respect to their neighbor and next nearest neighbor remains the same and therefore our discussion of the exchange interactions in section \ref{MagEI} does not depend on the definition of a unique spin direction.

\subsubsection{Magnetic ordering temperature of Ca$_3$LiOsO$_6$}

\begin{figure}[tb]
     \centering
                       \includegraphics[width=0.95\columnwidth]{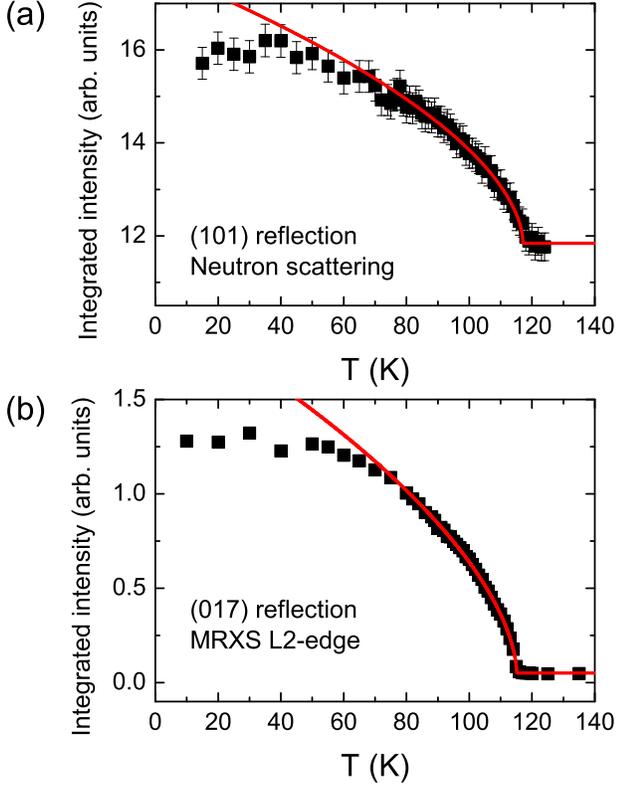}
 \caption{\label{FigureOP} Integrated intensity of the scattering at purely magnetic reflections with (a) powder neutron scattering and (b) single crystal MRXS. The determined antiferromagnetic transition ordering temperature $\rm T_{N}$ is shown to agree for both techniques.}
\end{figure}

Having experimentally determined the magnetic structure for Ca$_3$LiOsO$_6$ we now consider the antiferromagnetic ordering temperature, $\rm T_N$. We measured the scattering around the (101) reflection, $\rm |Q| = 0.97$ $\rm \AA^{-1}$, with elastic neutron scattering on beamline HB1 at HFIR and the scattering around the (107), $\rm |Q| = 4.16$ $\rm \AA^{-1}$, reflection at the L2-edge with MRXS on beamline 6-ID-B at the APS. The integrated intensity for the various temperature measurements are shown in Fig.~\ref{FigureOP}. We fit the temperature dependence of the integrated intensity to a power law to determine the magnetic ordering temperature and allow an estimate of the $\beta$ exponent. From the neutron scattering  results on a powder sample we find T$_{\rm N}=117.1 \pm 0.9$ K. The associated exponent is $\beta = 0.28 \pm 0.1$, however we stress that obtaining critical scattering from our data is not feasible and instead only note that the exponent is closer to the 3D value. We refer to the $(h,k,l)$ scans using MRXS as more direct evidence that the magnetic order is 3D (Fig.~\ref{FigureHKL_Xray}). MRXS gives a AFM ordering temperature of T$_{\rm N}\approx115$ K, however due to sample heating issues the reliability in sample temperature results in inassignable error bars and as such we simply note that the value is consistent with neutron scattering and bulk data \cite{ShiJACS}. The order parameter from both techniques has all the hallmarks for 3D order, and does not show the sharp increase associated with 2D ordering. Therefore within experimental error the neutron and x-ray results agree, and correspond to the transition temperature observed in susceptibly and specific heat measurements in the literature \cite{ShiJACS}.

\subsubsection{\label{MagEI}Magnetic exchange interactions}

\begin{figure}[tb]
     \centering
                       \includegraphics[width=0.9\columnwidth]{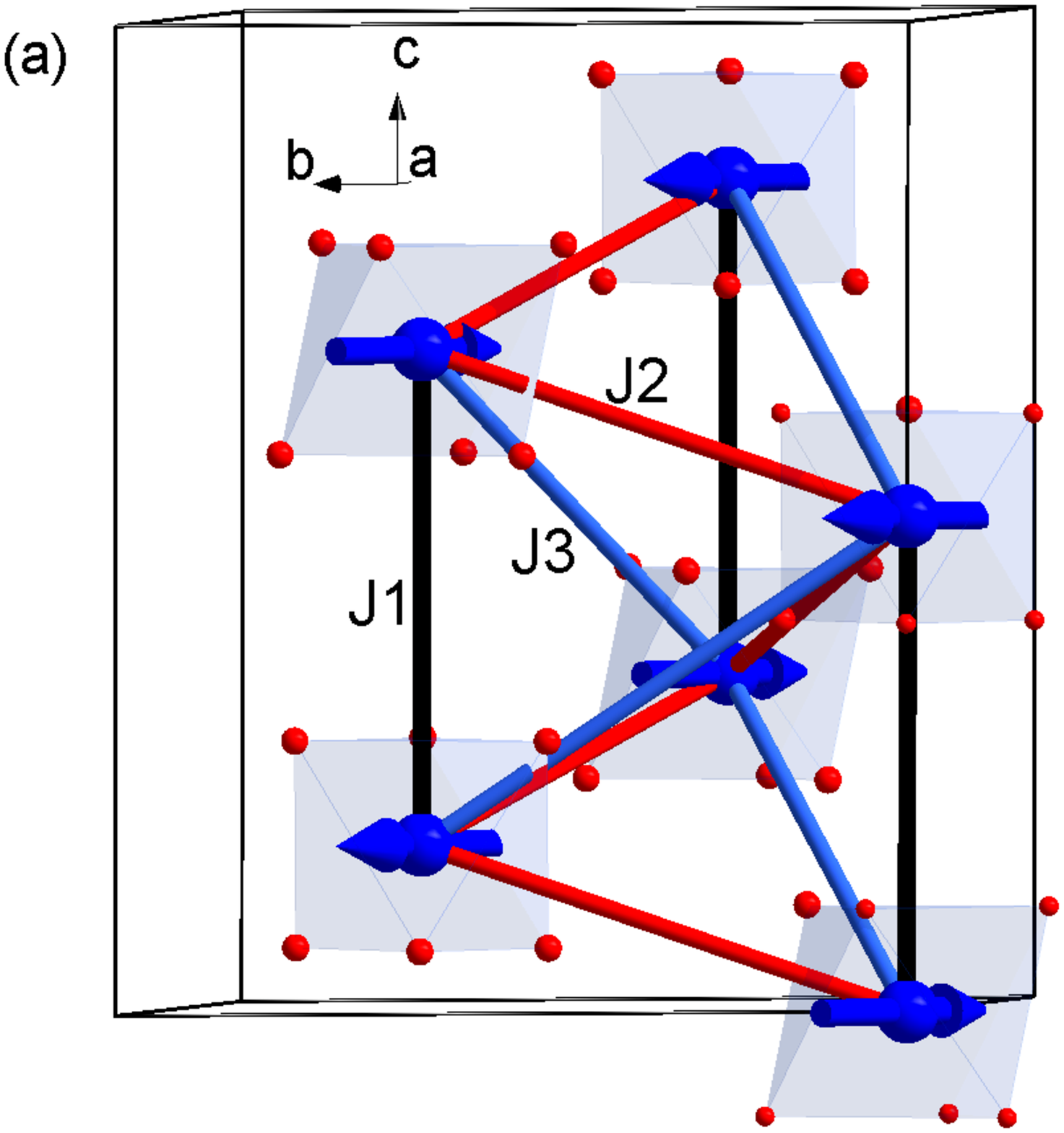} 
                                              \includegraphics[width=0.29\columnwidth]{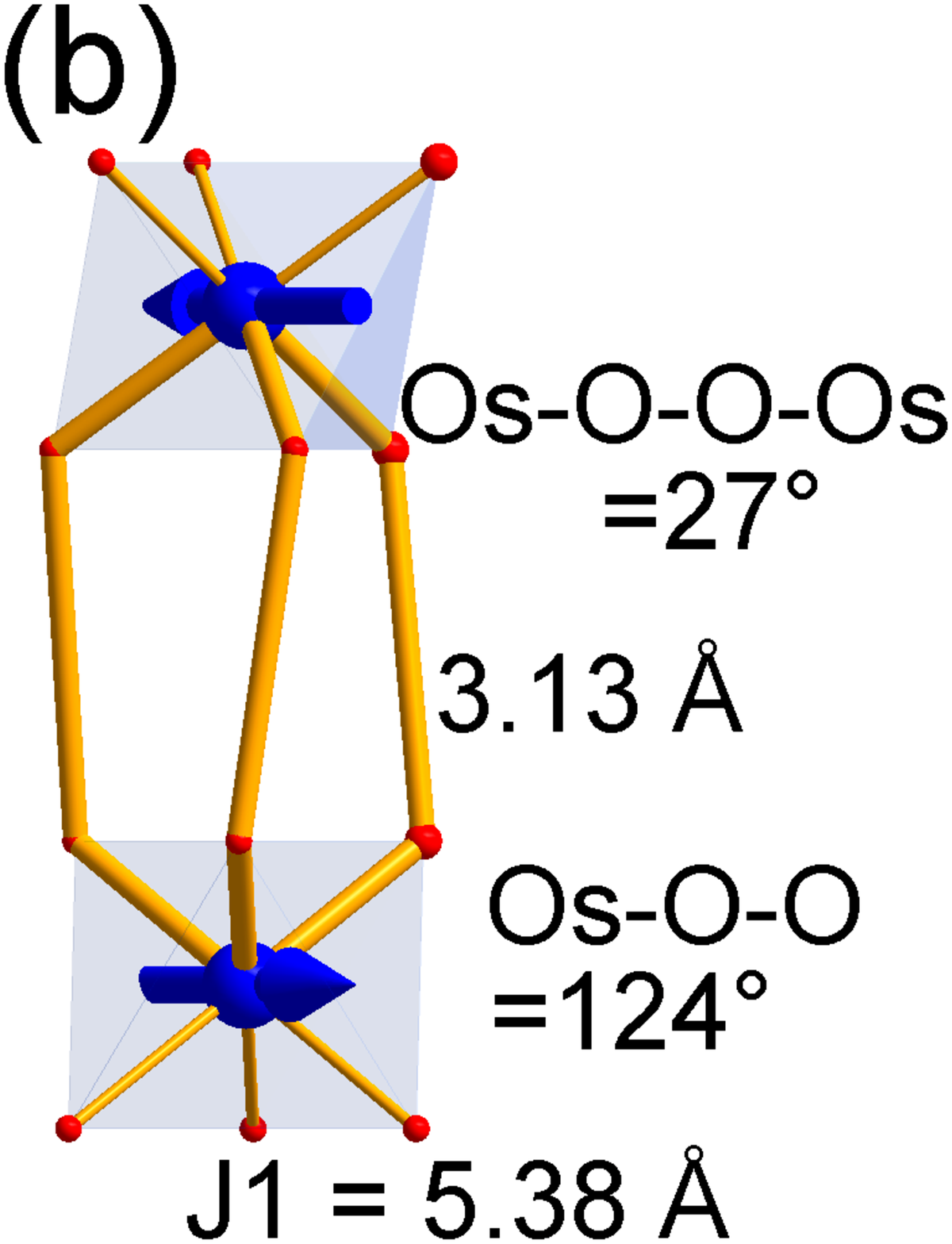} 
                                                                     \includegraphics[width=0.29\columnwidth]{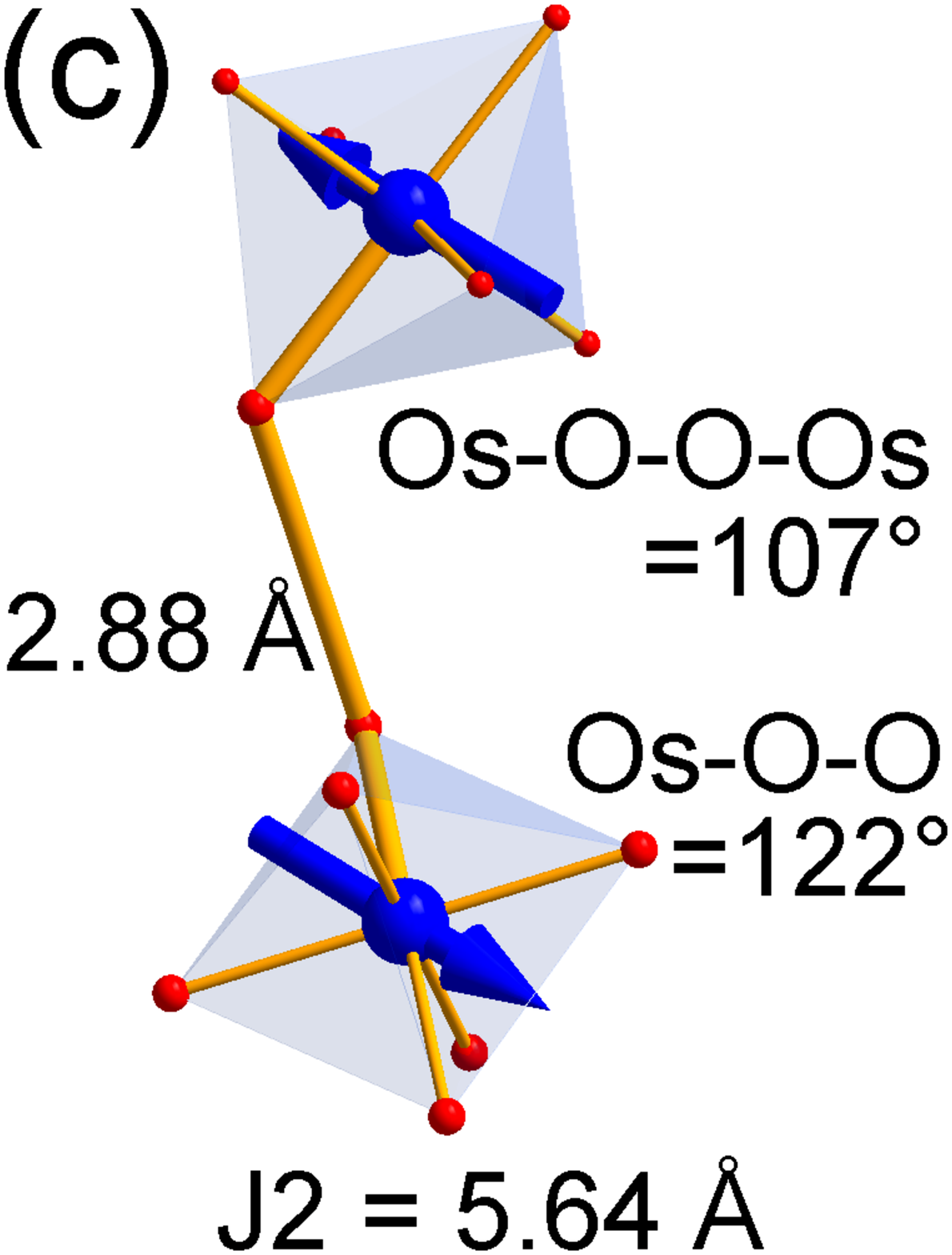} 
                                                                                            \includegraphics[width=0.29\columnwidth]{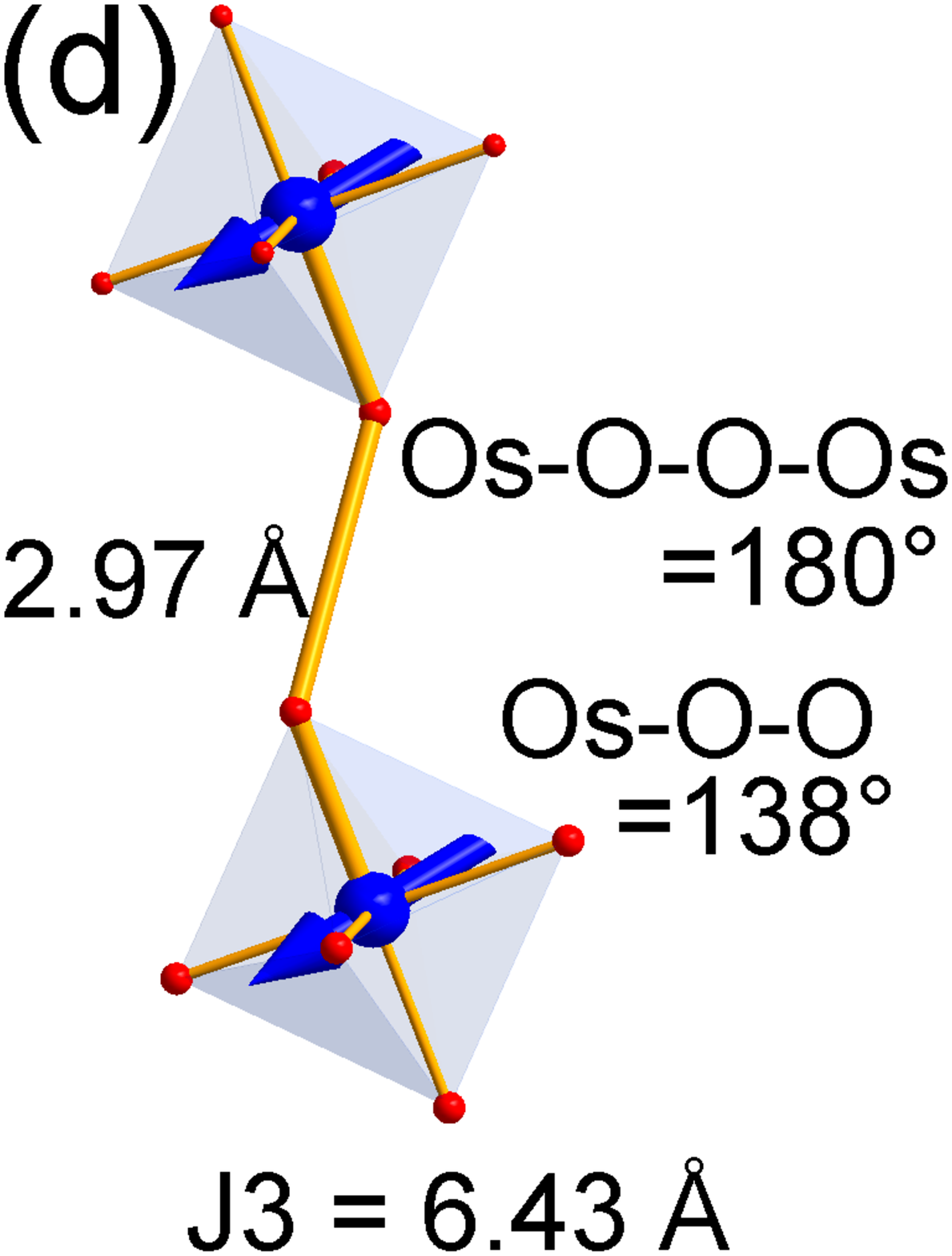} 
 \caption{\label{Jinteractions} Schematic of magnetic structure of Ca$_3$LiOsO$_6$. Blue spheres are Os ions with associated magnetic spin shown as an arrow. Smaller red spheres are oxygen ions. (a)  Magnetic extended superexchange pathways Os-O-O-Os: $J1$ (black), $J2$ (red) and $J3$ (blue). $J1$ and $J2$ form AFM exchange interactions, whereas $J3$ forms FM bonds. (b)-(d) The separate magnetic interaction pathways are shown with their respective distances and angles. The angle between spins in the schematic is 180$^{\circ}$. See the text for discussion of other angles, however we note that altering the angle between spins does not effect our discussion of the different exchange interactions.}
\end{figure}

The magnetic spins reside on the $a$-$b$ plane and form 3D magnetic order that is best explained as involving the extended superexchange magnetic interaction pathway Os-O-O-Os. The extended superexchange interactions are shown in Fig.~\ref{Jinteractions}, labelled $J1$, $J2$ and $J3$. Kan {\it et al.}~considered the relative strength of these exchange interactions using density functional theory calculations for Ca$_3$LiOsO$_6$ for various AFM or FM interactions \cite{KanInorChem}. Considering Fig.~\ref{Jinteractions}, the magnetic structure for Ca$_3$LiOsO$_6$ we have presented gives J1 spins oppositely aligned, J2 spins oppositely aligned and J3 spins coaligned. Kan {\it et al.}~considered four possible magnetic structures and concluded that all $J1$, $J2$ and $J3$ interactions were AFM, whereas the experimentally determined magnetic structure presented here has $J3$ spins parallel. The overall Os-Os bond length increases in going from J1 (5.38 $\rm \AA$) to J2 (5.64 $\rm \AA$) to J3 (6.43 $\rm \AA$). However, since the magnetic structure involves 3D interactions, the magnetic ordering cannot simply be described in terms of the shortest $J1$ interaction along the $c$-axis. Therefore the further exchange interactions $J2$ and $J3$ are required. Since frustration effects are negligible, this requires that $J2$ and $J3$ not be of a similar strength, as noted previously \cite{KanInorChem}. This would be compatible with $J1$ and $J2$ having large negative AFM values, while $J3$ is small and consequently is fixed according to the $J1$ and $J2$ interactions and produces apparent FM interactions, even though the $J3$ exchange interaction is negative. However, one alternative conclusion for $J3$ spins being parallel could be argued in terms of the bond angles between the three different Os-O-O-Os pathways favoring FM interactions, considered in Fig~\ref{Jinteractions}(b)-(d). For $J1$ and $J2$ the Os-O-O bond angle is approximately equivalent to 90$^\circ$ and the torsion angle of Os-O-O-Os deviates from 90$^\circ$ by approximately the same value. Comparing $J1$ and $J2$ with $J3$ shows markedly different Os-O-O and Os-O-O-Os bond angle values. Therefore the oxygen $p$ orbitals and Os $t_{2g}$ $d$ orbitals will overlap differently for $J3$, compared to $J1$ and $J2$. It would be of interest to theoretically calculate stable $J1$-AFM, $J2$-AFM and $J3$-FM values compatible with the magnetic structure for Ca$_3$LiOsO$_6$ we have reported.\\

\section{\label{sec:Conclusion}Conclusion}

We have presented a combined neutron and x-ray scattering investigation of Ca$_3$LiOsO$_6$. The results are compatible with a magnetic structure of $\Gamma(5)$ (following the numbering scheme of Kovalev) in which the spins are in the $a$-$b$ plane. The magnetic order is 3D and can be explained in terms of a model of Os-O-O-Os extended superexchange interactions. Despite apparent triangular units frustration is relieved as a consequence of the nature of the Os-O-O-Os pathways present that involve both AFM and FM exchange interactions. Ca$_3$LiOsO$_6$ provides an ideal model to investigate the magnetic extended superexchange interaction free from any single anion superexchange effects.

\begin{acknowledgments}
Work at ORNL was supported by the scientific User Facilities Division, Office of Basic Energy Sciences, U.~S.~Department of Energy (DOE). Use of the Advanced Photon Source, an Office of Science User Facility operated for the U.S. DOE Office of Science by Argonne National Laboratory, was supported by the U.S. DOE under Contract No. DE-AC02-06CH11357. Research was supported in part by Grant-in-Aid for Scientific Research (22246083, 22850019) from JSPS and FIRST Program from JSPS and ALCA program from JST and the Ministry of Science and Technology of China (973 Project No. 2011CBA00110).
\end{acknowledgments}

% Create the reference section using BibTeX:
%\bibliography{Ca3LiOsO6}

%merlin.mbs 2010-03-15 4.21a (PWD, AO, DPC)
%Control: key (0)
%Control: author (8) initials jnrlst
%Control: editor formatted (1) identically to author
%Control: production of article title (-1) disabled
%Control: page (0) single
%Control: year (1) truncated
%Control: production of eprint (0) enabled
%

\end{document}